\documentclass[aps,prx,amsmath,amssymb,citeautoscript,reprint]{revtex4-2}
\usepackage{graphicx, float, xcolor, pgf}
\usepackage{physics, siunitx}
\usepackage[caption=false, justification=centerlast]{subfig}

\makeatletter
\let\oldtheequation\theequation
\renewcommand\tagform@[1]{\maketag@@@{\ignorespaces#1\unskip\@@italiccorr}}
\renewcommand\theequation{(\oldtheequation)}
\makeatother

\usepackage{hyperref}
\hypersetup{
  pdfauthor={Amartya Bose},
  pdftitle={Effect of Phonons and Impurities on the Quantum Transport in XXZ Spin-Chains},
  colorlinks=true,
  linkcolor=black,
  citecolor=black,
  urlcolor=black
}

\begin{document}
\title{Effect of Phonons and Impurities on the Quantum Transport in XXZ Spin-Chains}
\author{Amartya Bose}
\thanks{Author to whom correspondence should be addressed}
\affiliation{Department of Chemistry, Princeton University, Princeton, New Jersey 08544, USA}
\email{amartyab@princeton.edu}
\email{amartya.bose@gmail.com}
\allowdisplaybreaks

\begin{abstract}
    Numerical and analytic results have been used to characterize quantum
    transport in spin chains, showing the existence of both ballistic and
    diffusive motion. Experiments have shown that heat transfer is surprisingly always diffusive. The scattering from phonons and
    impurities have been postulated to be the two factors critical in causing
    the diffusive transport. In this work, we evaluate the transport
    process by incorporating a bath of phonons and impurities in order to
    understand the role played by each of the factors. While methods like
    time-dependent density matrix renormalization group (tDMRG) can be used to
    simulate isolated spin chains, the coupling with phonons make simulations
    significantly more challenging. The recently developed multisite tensor
    network path integral (MS-TNPI) method builds a framework for simulating the
    dynamics in extended open quantum systems by combining ideas from tDMRG and
    Feynman-Vernon influence functional. This MS-TNPI is used to characterize
    dynamics in open, extended quantum systems. Simulations are done with the
    commonly used sub-Ohmic, Ohmic and super-Ohmic spectral densities describing
    the phononic bath. We show that while the transport in presence of
    impurities eventually becomes diffusive, the exact details are dependent on
    the specifics of the interactions and amount of impurities. In contrast, the
    presence of a bath makes the transport diffusive irrespective of the
    parameters characterizing the bath.
\end{abstract}
\maketitle

\section{Introduction}\label{sec:intro}
Non-equilibrium dynamics of quantum systems remain a major focus and important
challenge in
physics~\cite{gopalakrishnanAnomalousRelaxationHightemperature2019,prosenOpenXXZSpin2011,prosenExactNonequilibriumSteady2011,cornelissenLongdistanceTransportMagnon2015}.
Of special interest are the transport properties of spin chains. This is not
only because of possible applications in quantum information processing,
information storage, and spintronics, but also because spin chains provide us
with a relatively simple model, which can manifest complex aspects of
non-equilibrium dynamics. Experimental work done with ultra-cold
atoms~\cite{kuklovCounterflowSuperfluidityTwospecies2003,
    mandelCoherentTransportNeutral2003} has demonstrated the possibility of
representing these systems as spin-$\tfrac{1}{2}$ chains. The two states of a
spin represent atoms of different types occupying the given lattice site.

Recent work has studied the transport in quantum systems using a generalization
of hydrodynamics~\cite{castro-alvaredoEmergentHydrodynamicsIntegrable2016,
    bertiniTransportOutofequilibriumXXZ2016}, successfully predicting ballistic
current starting from inhomogenous initial states. However, such theories are
unable to predict the dynamics easy-axis regime. It is known that when the
systems have parity symmetries ($\mathbb{Z}_2$), and the initial state is
symmetric under the parity operation and under spatial reflection (ie.
$x\to-x$), the transport may lose the ballistic scaling when the observable is
odd under the parity~\cite{ljubotinaSpinDiffusionInhomogeneous2017}. This
implies that the ``rate'' of transport of the conserved quantity across the
``boundary'' at $q=0$ is sub-linear with time. Numerical simulations using
time-dependent density matrix renormalization group
(tDMRG)~\cite{whiteRealTimeEvolutionUsing2004,
    schollwockDensitymatrixRenormalizationGroup2011a,
    vidalEfficientSimulationOneDimensional2004, paeckelTimeevolutionMethodsMatrixproduct2019} have been used to explore this
regime of quantum transport in XXZ chains~\cite{gobertRealtimeDynamicsSpin2005,
    ljubotinaSpinDiffusionInhomogeneous2017}, demonstrating the superdiffusive
dynamics for the isotropic spin chain.

The XXZ-model has proved to be extremely useful, not just for the theoretical
study of transport properties~\cite{gobertRealtimeDynamicsSpin2005,
    ljubotinaSpinDiffusionInhomogeneous2017,
    bertiniTransportOutofequilibriumXXZ2016}, but also in describing real
systems~\cite{hlubekSpinonHeatTransport2012}. Despite the predictions of
ballistic heat transport in these systems,
\citet{hlubekSpinonHeatTransport2012} observed that the transport was
diffusive. This was attributed to extrinsic scattering of the spinons off
impurities and phonons. It is thus important to understand the contribution of
each of these scattering events in bringing about the diffusive transport.

The simulations of spin chains with impurities are relatively simple. One can
simulate a statistical ensemble of distribution of impurities in spin chains
using tDMRG or the time-dependent variational principle (TDVP)~\cite{haegemanTimeDependentVariationalPrinciple2011,haegemanUnifyingTimeEvolution2016, klossTimedependentVariationalPrinciple2018, gotoPerformanceTimedependentVariational2019}. However, the incorporation of coupling to phonons in the dynamics
of extended quantum systems at a finite temperature is computationally extremely
challenging. The presence of low frequency modes at high equivalent temperatures
neccessitate the use of large bases for the phononic bath and consequently leads
to an exponential growth of computational complexity for wave function-based
methods. Some interesting work has been done to understand the dynamics of
boundary-driven XXZ
chain~\cite{prosenExactNonequilibriumSteady2011,prosenOpenXXZSpin2011}.

Open quantum systems are most often simulated by integrating out the phononic
bath using path integrals based on the Feynman-Vernon influence
functional~\cite{feynmanTheoryGeneralQuantum1963}. It has been shown that it is
possible to enhance the performance of influence functional simulations using
tensor network~\cite{strathearnEfficientNonMarkovianQuantum2018,
    jorgensenExploitingCausalTensor2019, boseTensorNetworkRepresentation2021,
    bosePairwiseConnectedTensor2022}. Despite such advances, the simultaneous
presence of an extended system and the phonons leads to problems that cannot be
easily simulated. The existence of local baths introduces a non-Markovian
memory, within which the scaling of the computational requirements scale
exponentially. For extended systems the base of the scaling is so large that the
computations are infeasible even for small memory lengths. We have developed an
extension to the framework of
tDMRG incorporating influence functionals leading to a method called multisite
tensor network path integral
(MS-TNPI)~\cite{boseMultisiteDecompositionTensor2022}.  MS-TNPI is able to
simulate the dynamics of open, extended quantum systems accounting for
non-Markovian memory. This has been used advantageously to simulate the
excitonic dynamics in photosynthetic complexes~\cite{boseTensorNetworkPath2022}.

The primary objective of this paper is to numerically explore the role played
by the scattering from phonons and impurities in the non-equilibrium transport
in antiferromagnetic XXZ chain. In Sec.~\ref{sec:system}, the systems explored
are described, along with some well-known properties. Standard tDMRG is used
for simulating the transport process in presence of impurities. MS-TNPI is used
when accounting for the non-Markovian coupling to phonons and is described in
Sec.~\ref{sec:mstnpi}. The results of the simulation are presented in
Sec.~\ref{sec:results}. We will show that the presence of phonons causes the
dynamics to become diffusive, irrespective of the characteristics of the
phononic modes. The presence of impurities on the other hand changes the
dynamics in more subtle ways, that are dependent on the particulars of the
interactions introduced by the impurities. Therefore, the observation of
uniformly diffusive dynamics in Ref.~\cite{hlubekSpinonHeatTransport2012} is likely
because of interactions with phonons. We end with some concluding remarks and future prospects in Sec.~\ref{sec:conclusion}.

\section{Systems under study}\label{sec:system}
Consider an XXZ spin chain with $n$ spins (for even $n$). The Hamiltonian is
given by
\begin{align}
    \hat{H}_0 & = \hbar J \sum_{-\frac{n}{2}<q<\frac{n}{2}}\left(\hat s_q^{(1)}\hat s_{q+1}^{(1)}+\hat s_q^{(2)}\hat s_{q+1}^{(2)}+\Delta\hat s_q^{(3)}\hat s_{q+1}^{(3)}\right)\label{eq:XXZ}
\end{align}
In Eq.~\ref{eq:XXZ}, $\hat{s}_q^{(k)}$ are the spin-$\frac{1}{2}$ operators for
the spin at the spatial location $q$. It is related to the Pauli matrices as
$\hat{s}_q^{(k)} = \frac{1}{2}\hat\sigma_q^{(k)}$. The total magnetization $M =
    \sum_q \hat{s}_q^{(3)}$ is a conserved quantity in the XXZ model. The anisotropy
in the system is encoded in $\Delta$. If $\Delta = 0$, the XXZ model reduces to
the Frenkel problem, which ubiquitous in the study of exciton transfer. The two
eigenstates of $\hat\sigma_q^{(3)}$ with eigenvalues of $\pm 1$ are denoted as
$\ket{\uparrow_q}$ and $\ket{\downarrow_q}$ respectively, with the $q$ subscript
omitted where it does not cause ambiguity. Depending on the sign of $\Delta$,
the ground state is either ferromagnetic ($\Delta<0$) or antiferromagnetic
($\Delta>0$). Here we consider the antiferromagnetic system with $\Delta>0$. The
excitation spectrum is gapped for $|\Delta|>1$, and when $|\Delta|<1$, the
system becomes gapless and the correlation functions show power law
behavior~\cite{gobertRealtimeDynamicsSpin2005}. The case of $\Delta=0$ leads to
the so-called XX model, also known in the literature as the Frenkel model of
exciton transfer.

In this paper, the non-equilibrium transport is studied from the initial state
\begin{align}
    \tilde\rho(0) & = \dyad{\uparrow\uparrow\ldots\uparrow\downarrow\ldots\downarrow\downarrow},
\end{align}
where the left half spins are in up state ($\ket{\uparrow}$) and the spins in
the right half of the chain are in down state ($\ket{\downarrow}$). This fully
polarized domain wall is a very special setup that has received attention in the
literature~\cite{gobertRealtimeDynamicsSpin2005}. In other explorations an
initial state with a more ``tilted'' domain wall is
selected~\cite{ljubotinaClassStatesSupporting2017,ljubotinaSpinDiffusionInhomogeneous2017}.
The exact nature of the non-equilibrium dynamics is sensitive to the exact
initial condition.  The transport process can be characterized quantitatively
through the scaling of the time-dependent spin profile and the magnetization
transferred between the two halves given as an integral over the spin current,
$j$, at $q=0$:
\begin{align}
    \Delta m & = \int_0^t j(0, t') \dd{t'}
    \\ & \propto \sum_{q>0} \left(\hat{s}_q^{(3)}(t) + \frac{1}{2}\right)  \propto t^\alpha.
\end{align}
A scaling of $\alpha = 1$ implies ballistic motion, and $\alpha = 0.5$ implies
diffusive motion. Superdiffusive motion is characterized by $0.5<\alpha<1$.

The first and computationally simpler mechanism for the change of the nature of
transport that was postulated and is explored here is scattering from
impurities. Previous work~\cite{znidaricWeakIntegrabilityBreaking2020} has shown
that addition of even an infinitesimal periodic perturbation can make the
thermodynamic limit transport diffusive. Here, we take a different approach to
modeling impurities in a spin chain. We assume that an impurity is distinguished
from a usual site by the way it interacts with its neighboring spins. Addition
of a new type of site introduces two additional types of coupling: the
impurity-site and the impurity-impurity couplings. In this work, we take the
differences to be in the value of $\Delta$. Different random configurations with
certain proportions of impurities are sampled over to obtain the average
dynamics. This is explored in Sec.~\ref{sec:impurities}.

Subsequently we explore the effect of scattering from phonons in
Sec.~\ref{sec:phonons}. In presence of phonons, the Hamiltonian is modified
through interactions with the harmonic bath describing the phonons,
\begin{align}
    \hat{H} & = \hat{H}_0 + \hat{H}_\text{spin-phonon},
\end{align}
where $\hat{H}_0$ is the Hamiltonian of the isolated XXZ spin-chain,
Eq.~\ref{eq:XXZ}, and $\hat{H}_\text{spin-phonon}$ is the Hamiltonian
corresponding to the dissipative phononic bath interacting with the system at
$x$.
\begin{align}
    \hat{H}_\text{spin-phonon} & = \sum_{-\frac{n}{2}<q<\frac{n}{2}}\sum_{j} \frac{p_{j,q}^2}{2 m_{j,q}}\nonumber                                     \\
                               & + \frac{1}{2}m_{j,q}\omega_{j,q}^2 \left(x_{j,q} - \frac{c_{j,q} \dyad{\uparrow_q}}{m_{j,q}\omega_{j,q}^2}\right)^2,
\end{align}
where $\omega_{j,q}$ and $c_{j,q}$ are the frequency and coupling of the
$j$\textsuperscript{th} mode of site $q$. The interaction between the spin
chain and the phonons is such that the harmonic oscillators of the phonon get
shifted only when the corresponding spin is in the $\ket{\uparrow}$ state. The
bath is usually characterized by a spectral density,
\begin{align}
    J(\omega) = \frac{\pi}{2}\sum_j \frac{c_j^2}{m_j\omega_j}\delta(\omega - \omega_j).
\end{align}
One of the most important characteristics of a harmonic bath is the
reorganization energy, $\lambda=\frac{1}{\pi}\int_{0}^\infty \dd{\omega}
    \frac{J(\omega)}{\omega}$. For the current study, we use the well-known form with an exponential cutoff:
\begin{align}
    J(\omega) = 2\pi\hbar\xi\frac{\omega^s}{\omega_c^{s-1}}\exp\left(-\frac{\omega}{\omega_c}\right),
\end{align}
where $\omega_c$ is the cutoff frequency and $\xi$ is the dimensionless Kondo
parameter encoding the strength of spin-bath coupling. The type of the bath is
determined by the value of $s$: $s<1$ defines a sub-Ohmic bath, $s=1$ defines an
Ohmic bath, and $s>1$ is a super-Ohmic bath. The reorganization energy of the
bath is $\lambda = 2\hbar\omega_c\xi\Gamma(s)$.

For the simulations with interaction with phonons, the initial state is taken to be a
product state between the reduced density matrix of the spin chain and the
thermal density of the isolated bath:
\begin{align}
    \rho(0) & = \tilde\rho(0) \otimes \frac{\exp(-\beta\hat{H}_\text{phonon})}{Z_\text{phonon}}.
\end{align}
Here $Z_\text{phonon}$ is the partition function for the bath at an inverse
temperature of $\beta = \frac{1}{k_BT}$.

The simulation of transport in presence of phonons at a non-zero temperature is
challenging because of the presence of temporally non-local interactions in the
form of non-Markovian memory. The recently introduced MS-TNPI
method~\cite{boseMultisiteDecompositionTensor2022} allows us to capture these
non-Markovian effects in a numerically exact Feynman-Vernon influence
functional-based formalism using tensor networks. This method is described in
short in Sec.~\ref{sec:mstnpi}.

\section{Multisite Tensor Network Path Integral}\label{sec:mstnpi}

While tDMRG is well-suited for the simulation of extended quantum systems like
spin chains, the presence of dissipative media poses significant computational
challenges. The recently developed multisite tensor network path integral
(MS-TNPI)~\cite{boseMultisiteDecompositionTensor2022} extends tDMRG ideas to
account for presence of harmonic modes. If the initial state can be expressed as
a direct product of the system's initial reduced density matrix and the bath's
thermal density, then the time-propagated reduced density matrix of the system
is given as
\begin{widetext}
    \begin{align}
        \mel{S_N^+}{\tilde\rho(N\Delta t)}{S_N^-} & = \sum_{S_0^\pm}\sum_{S_1^\pm}\ldots\sum_{S_{N-1}^\pm} \bra{S_N^+}\hat{U}\dyad{S_{N-1}^+}\hat{U}\ket{S_{N-2}^+}\ldots\nonumber                           \\
                                                  & \times\bra{S_1^+}\hat{U}\dyad{S_0^+}\tilde\rho(0)\dyad{S_0^-}\hat{U}^\dag\ket{S_1^-}\ldots\bra{S_{N-1}^-}\hat{U}\ket{S_N^-} F[\{S^\pm_j\}] \label{eq:pi}
    \end{align}
\end{widetext}
Here, $U$ is the short-time propagator for the spin chain, $S_j^\pm$ represents the
``forward-backward'' state of the spin chain at the $j$\textsuperscript{th}
time-point. In the spin chain with many sites, $S_j^\pm$ is a short-hand for
$s_{i,j}^\pm$ where the first index, $i$, is the index of the spatial location
of the site, $q$, and the second index, $j$ gives the time point. The
Feynman-Vernon influence functional~\cite{feynmanTheoryGeneralQuantum1963},
denoted by $F[\{S^\pm_j\}]$, is dependent upon the history of the system. The
baths are assumed to be site local. Therefore the total influence functional is
a product of the individual influence functionals corresponding to each of the
sites:
\begin{align}
    F[\{S^\pm_j\}] & = \prod_{i}\exp\left(-\frac{1}{\hbar}\sum_{k=0}^{N}\Delta s_{i,k}\sum_{k'=0}^{k}(\eta^{(i)}_{kk'}s_{i,k'}^+ - \eta^{(i)*}_{kk'}s_{i,k'}^-)\right),\label{eq:fvif}
\end{align}
where $\Delta s_{i,k} = s^+_{i,k} - s^-_{i,k}$ and $\eta^{(i)}_{kk'}$ are the
discretized influence functional
coefficients~\cite{makriTensorPropagatorIterativeI1995,makriTensorPropagatorIterativeII1995}.
These $\eta$-coefficients are generally expressed as integrals involving the
spectral density.

There are two basic entities required for simualting the path integral: (1) a
forward-backward propagator $K(S^\pm_j, S^\pm_{j+1}) =
    \mel{S_{j+1}^+}{\hat{U}}{S_j^+}\mel{S_j^-}{\hat{U}^\dag}{S_{j+1}^-}$, and (2)
the influence functional corresponding to each path. For an extended spin chain,
the cost of storing the full propagator becomes prohibitive. Thus, following the
time-evolving block decimation (TEBD) method for propagation of the
wavefunction~\cite{paeckelTimeevolutionMethodsMatrixproduct2019}, the ``system
axis,'' $S^\pm_j$ is also factorized out, yielding a matrix product
representation of the forward-backward propagator. In TEBD, this propagator is
repeatedly applied to the state to simulate time propagation. This dynamics is
Markovian.

\begin{figure}
    \includegraphics[scale=0.30]{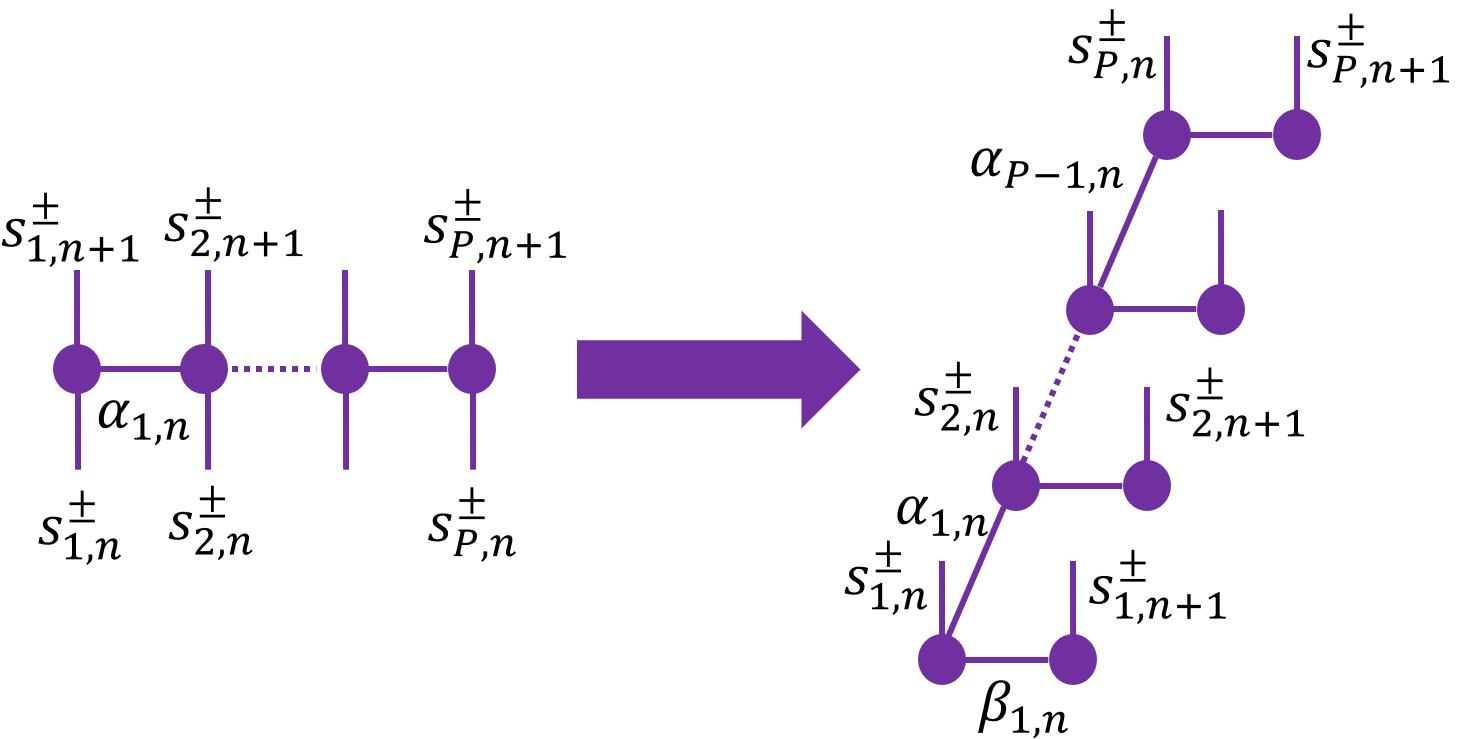}
    \caption{Refactorization of the propagator MPO.}\label{fig:prop_refactor}
\end{figure}

\begin{figure}
    \subfloat[MS-TNPI network]{\includegraphics[scale=0.2]{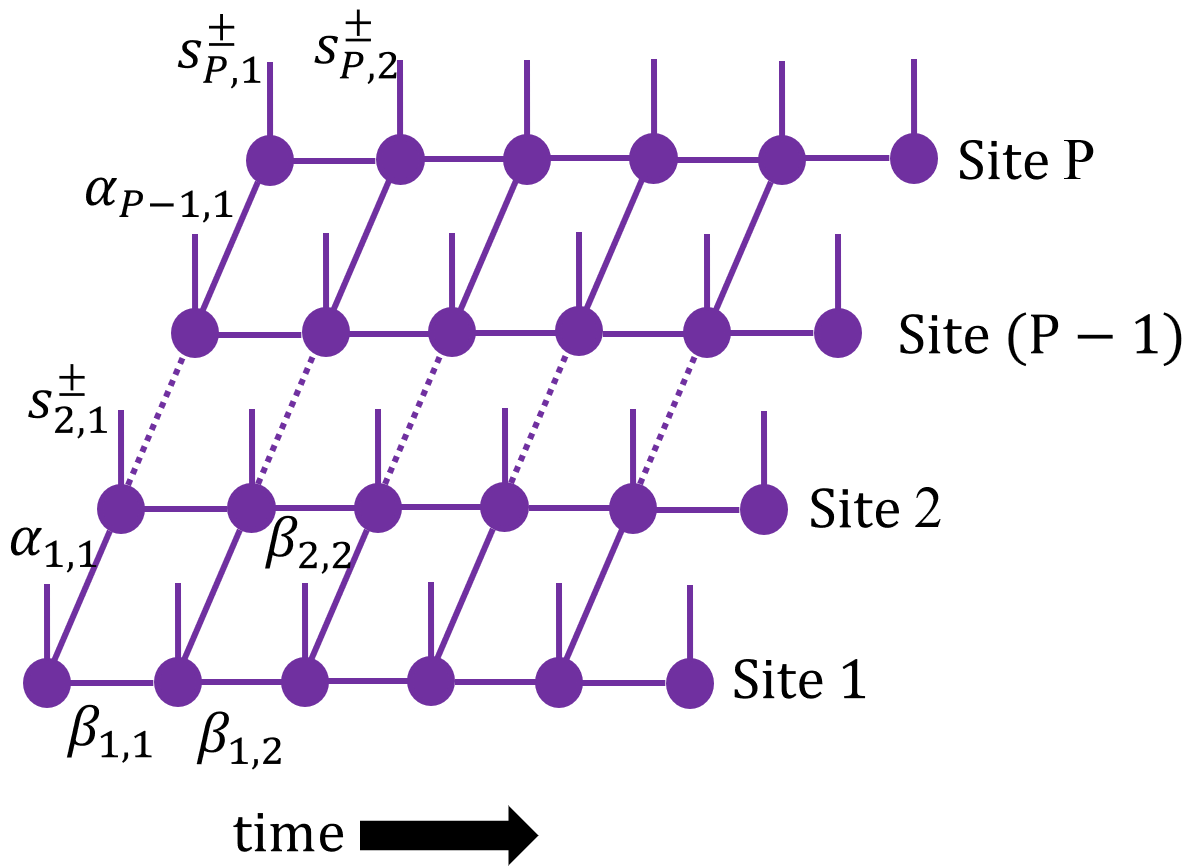}}
    ~\subfloat[Influence functional incorporation]{\includegraphics[scale=0.2]{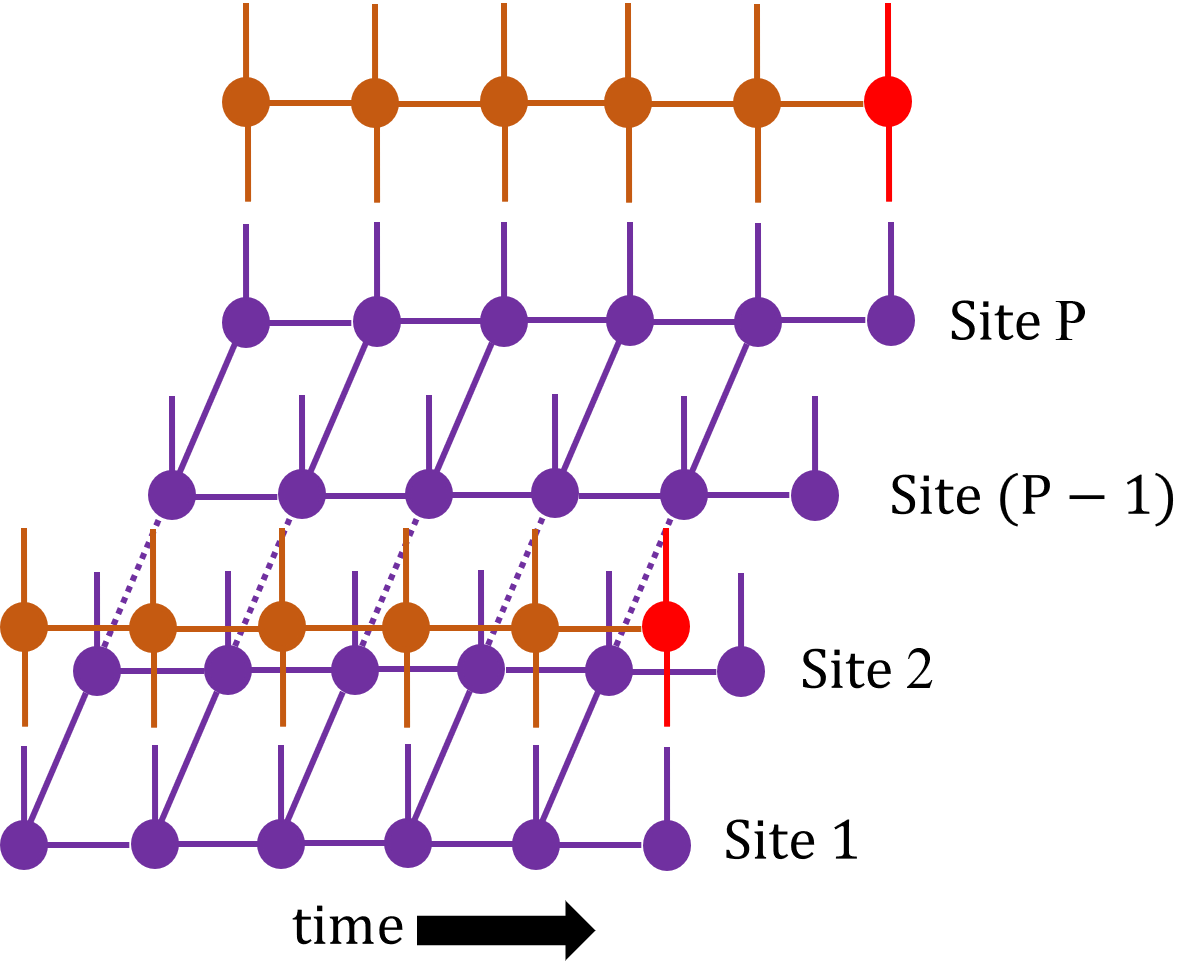}}
    \caption{MS-TNPI network obtained by multiplying the refactorized propagator MPOs.}\label{fig:mstnpi}
\end{figure}

Note that the repeated application of the propagator MPO to the state MPS
involves an automatic summation over the ``previous'' system state and yields
the propagated state MPS. The process of incorporating the impact of the bath
through the Feynman-Vernon influence functional neccessitates the preservation
of the state of the extended quantum system over the length of history. This is
not possible in the MPO-MPS propagation scheme.  Alternatively, one could
imagine multiplying the different propagator MPOs in a direct product sense,
which would leads to storage of exponentially large tensors. MS-TNPI solves this
problem by factorizing the forward-backward propagator matrix product operator
(MPO) as shown in Fig.~\ref{fig:prop_refactor}. This separation of the
``initial'' index and the ``final'' index into different tensors allow us to
assemble multiple time points in the form of a two-dimensional tensor network
through ``direct products'' over the individual sites. This network retains the
information of the non-Markovian history is schematically shown in
Fig.~\ref{fig:mstnpi}~(a).

In this 2D MS-TNPI tensor network, Fig.~\ref{fig:mstnpi}~(a), the rows represent the
path amplitude tensors corresponding to each of the units, and the column
corresponds to the different time points in history. If one were to contract the
network along the rows accumulating the columns, the method would be equivalent
to a density matrix extension of TEBD or tDMRG. However, the 2D structure allows
the incorporation of the influence functional in the form of matrix product
operators. Because the baths are site-local, these MPOs act along the rows. This
is schematically indicated in Fig.~\ref{fig:mstnpi}~(b). Note that for most
baths representing condensed phase dissipative environments, the memory length
is not infinite. It can be truncated, and the dynamics can be numerically
converged with respect to this memory length. The algorithm for doing this
finite memory iteration is outlined in detail in
Ref.~\cite{boseMultisiteDecompositionTensor2022}.


\section{Results}\label{sec:results}
\subsection{Transport in presence of impurities}\label{sec:impurities}
We start the study by analyzing the transport in an XXZ chain with impurities.
These simulations are done with standard TEBD. In Fig.~\ref{fig:isolatedXXZ}, we
demonstrate the quantum transport in the isolated XXZ spin chain. As is
well-known~\cite{gobertRealtimeDynamicsSpin2005}, the transport is ballistic for
$0 < \Delta < 1$ ($\alpha=1$), diffusive for $\Delta > 1$ ($\alpha=0.5$, not
simulated here), and superdiffusive for $\Delta = 1$ ($\alpha\approx\frac{3}{5}$).
While $\alpha\approx\frac{3}{5}$ for the fully polarized domain boundary simulated
here~\cite{gobertRealtimeDynamicsSpin2005,
    ljubotinaSpinDiffusionInhomogeneous2017}, the exact scaling for the isotropic
XXZ model is dependent on the initial condition used. In particular,
$\alpha\approx\frac{2}{3}$ at high
temperatures~\cite{ljubotinaSpinDiffusionInhomogeneous2017,znidaricSpinTransportOnedimensional2011}.
Figure~\ref{fig:isolatedXXZ}~(b) shows striations with certain wavefronts
showing ballistic motion. This seems to be qualitatively similar to what was
observed in Ref.~\cite{ljubotinaClassStatesSupporting2017}.

\begin{widetext}
    \begin{minipage}{\textwidth}
        \begin{figure}[H]
            \centering
            \subfloat[$\Delta = 0$. Ballistic transport. Guide line shows $q\sim t$.]{\includegraphics{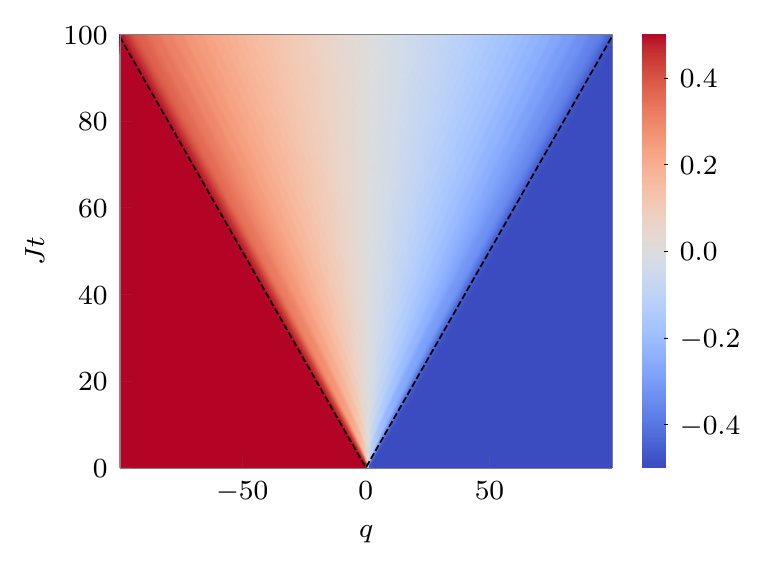}}
            ~\subfloat[$\Delta = 1$. Superdiffusive transport. Guide line shows $q\sim t^{3/5}$.]{\includegraphics{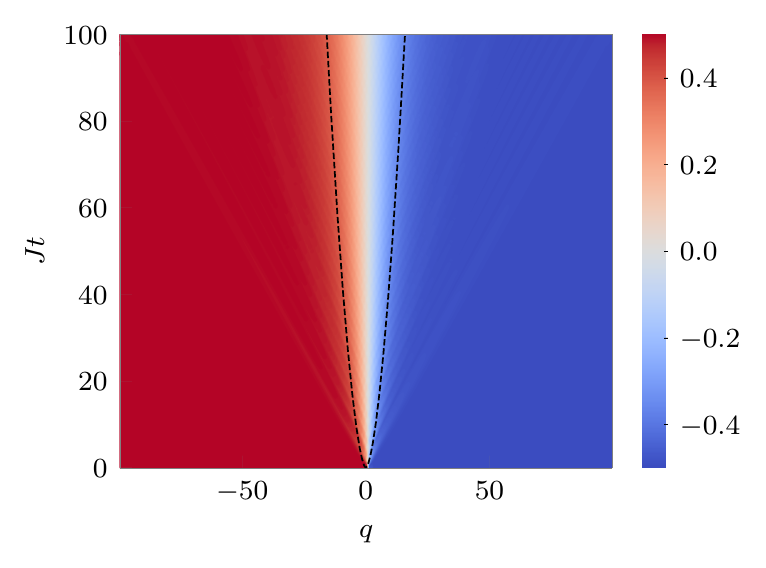}}
            \caption{Dynamics of $\expval{\hat{s}_q^{(3)}(t)}$ for the isolated XXZ spin chain at different values of anisotropy.}\label{fig:isolatedXXZ}
        \end{figure}
    \end{minipage}
\end{widetext}

Now, we start introducing impurities in the XXZ chain. The impurities considered
in this paper can be thought of as replacing the original sites with a different
kind of site. Suppose the base pure chain is taken to have an anisotropy of
$\Delta = 0$. For every site that has been replaced by an impurity, the
interaction with the neighboring sites changes depending on the nature of the
neighboring site.  For simplicity of simulation, we assume that the value of $J$
remains constant irrespective of whether the interaction is site-site,
site-impurity or impurity-impurity. However the anisotropy values are taken to
be different. For this preliminary exploration, we assumed
\begin{align}
    \Delta_\text{spin-spin}         & = 0,   \\
    \Delta_\text{spin-impurity}     & = 0.5, \\
    \Delta_\text{impurity-impurity} & = 1.
\end{align}
(Note that in this case, the pure spin chain would show ballistic dynamics, and
a spin chain made of 100\% impurities should show a super-diffusive dynamics.)
To study the effect of impurities, we simulate the dynamics at different
proportions of doping. Every site has a certain probability of being an
impurity, and the Hamiltonian is consequently defined by the arrangement of
these impurities. The overall dynamics is obtained as a statistical average over
such arrangements.

\begin{figure}
    \subfloat[50\% probability of impurity]{\includegraphics{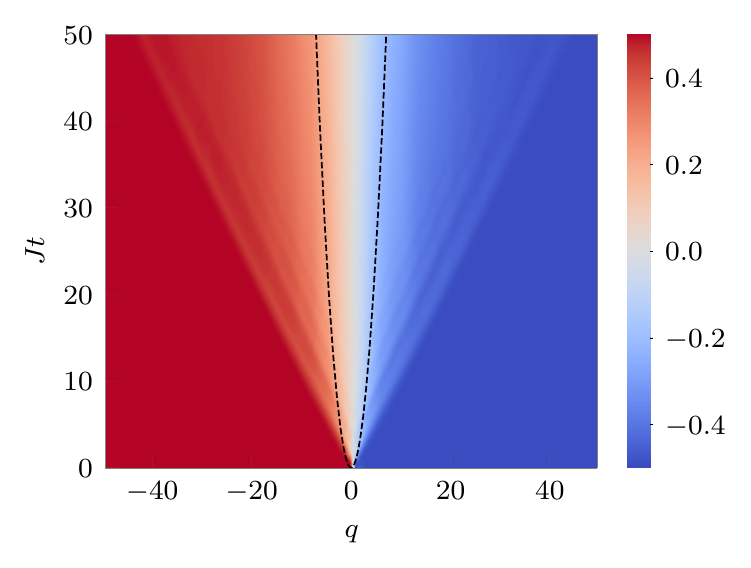}}

    \subfloat[75\% probability of impurity]{\includegraphics{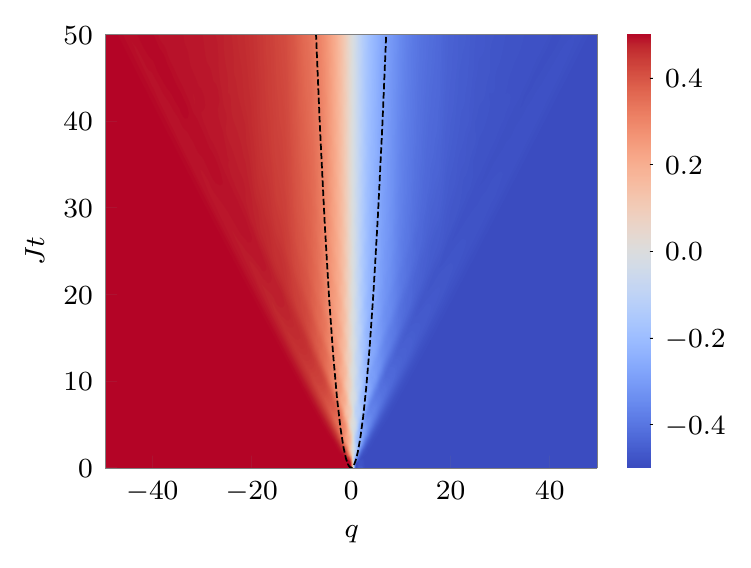}}
    \caption{Dynamics of $\expval{\hat{s}_q^{(3)}(t)}$ for an XXZ chain with different probabilities of each site being an impurity with dashed guide for the
    $\alpha = 0.5$ scaling.}\label{fig:impure}
\end{figure}

\begin{figure}
    \includegraphics{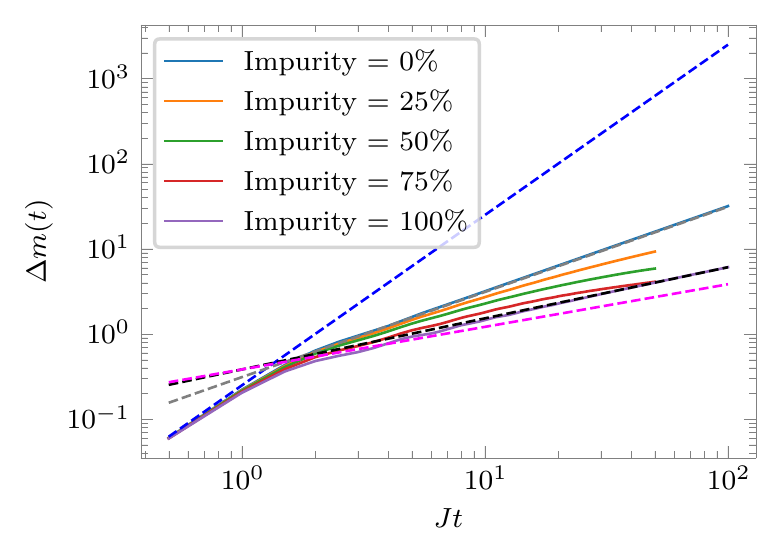}
    \caption{Transfer of magnetization as a function of time for different levels of impurity. Blue dashed line: $\Delta m \sim t^2$. Gray dashed line: $\Delta m \sim t$. Black dashed line: $\Delta m \sim t^{3/5}$. Magenta dashed line: $\Delta m \sim t^{1/2}$.}\label{fig:dope_mag}
\end{figure}

First, in Fig.~\ref{fig:impure} let us consider the dynamics of
$\expval{\hat{s}_q^{(3)}(t)}$ for the XXZ spin chain with the sites having 25\%
probability of being an impurity. The amount of magnetization transferred across
the initial domain boundary at $q = 0$ is shown as a function of time in
Fig.~\ref{fig:dope_mag} for different percentages of doping. First, notice that
the extremely short time dynamics is ``super-ballistic,'' with $\alpha = 2$. At
intermediate time-scales, the dynamics shows a continuous gradation of the
effective scaling with the probability of impurity. While the simulations have
not been run for long enough durations, the ones with a 75\% probability of
impurity seems to show a distinct slowing down of dynamics to diffusive
transport. This is in line with the observations of long time dissipative
dynamics in presence of perturbations that make the system
non-integrable~\cite{znidaricWeakIntegrabilityBreaking2020}.

Therefore, the numerical evidence seems to indicate that although the presence
of impurities make the dynamics diffusive at long times, there are non-diffusive
transients that reflect the exact nature of the impurities. These transients
last for longer timespans when the amount of impurity is less. A more thorough
exploration of the dynamics, while interesting, is beyond the scope of this
work. This study will be a topic of future work.

\subsection{Transport in presence of phonons}\label{sec:phonons}

\begin{figure*}
    \subfloat[Dynamics of $\expval{\hat{s}_q^{(3)}(t)}$.]{\includegraphics{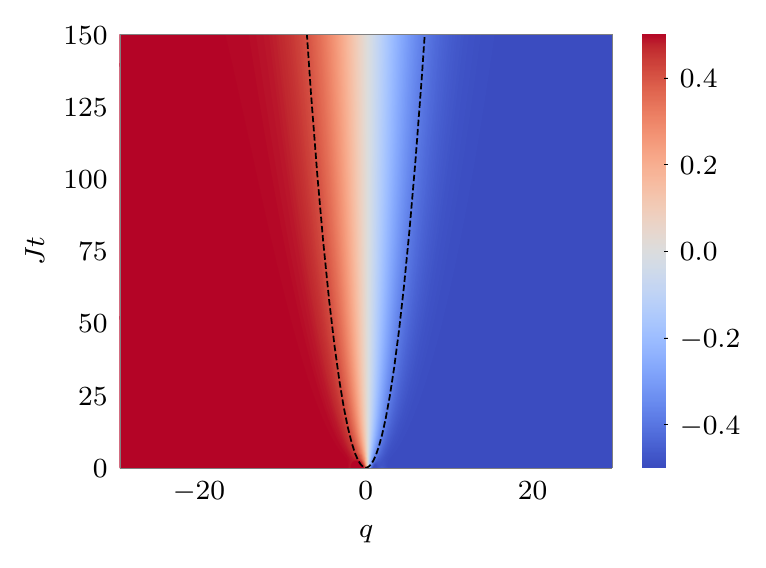}}
    ~\subfloat[Expectation values of $\hat{s}_q^{(3)}$ for $\Delta = 0$ at different time points]{\includegraphics{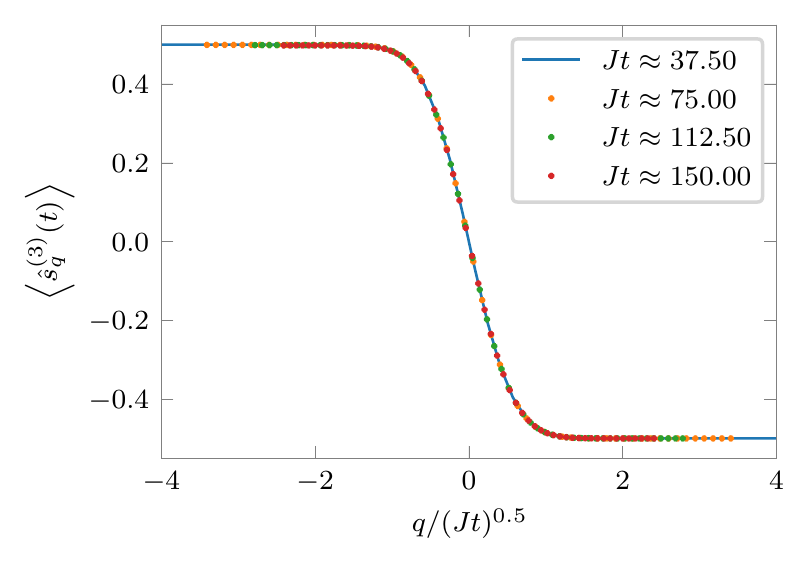}}
    \caption{Dynamics of $\expval{\hat{s}_q^{(3)}(t)}$ for an XXZ spin chain with $\Delta = 0$ in presence of an Ohmic bath ($s=1$) characterized by $\omega_c = 10J$, $\xi = 1$, and $\hbar\omega_c\beta = 5$. The black dashed lines guide the eye towards a scaling of $q\sim\sqrt{t}$.}\label{fig:spin_dyn_D0}
\end{figure*}

Addition of a dissipative medium (phonons, in this case) to each of the sites
``smears'' out the dynamics. First, we explore the dynamics of
$\expval{\hat{s}_q^{(3)}(t)}$ as a function of time in
Fig.~\ref{fig:spin_dyn_D0} for $\Delta = 0$.  In this case we consider a
relatively strongly coupled, cold and fast Ohmic ($s=1$) bath with $\omega_c =
    10J$, $\xi = 1$ and $\hbar\omega_c\beta = 5$. Also shown along with the dynamics
as a function of both time and site location, we also report the spin profile
various sites at different times. From Fig.~\ref{fig:spin_dyn_D0}~(b), it is
clear that the spin profile is invariant under a scaling of $q\sim\sqrt{t}$.
Thus we have demonstrated that the presence of the Ohmic bath converted the
ballistic dynamics of the isolated XXZ chain with $\Delta = 0$ to a diffusive
dynamics. Note that the ballistic wavefronts observed in
Fig.~\ref{fig:isolatedXXZ} and Fig.~\ref{fig:impure} are completely absent in
presence of the phonons in Fig.~\ref{fig:spin_dyn_D0}. (The graphs for $\Delta =
    1$ and $\Delta = 2$ are shown in Appendix~\ref{app:dynamics}.)

\begin{figure}
    \includegraphics{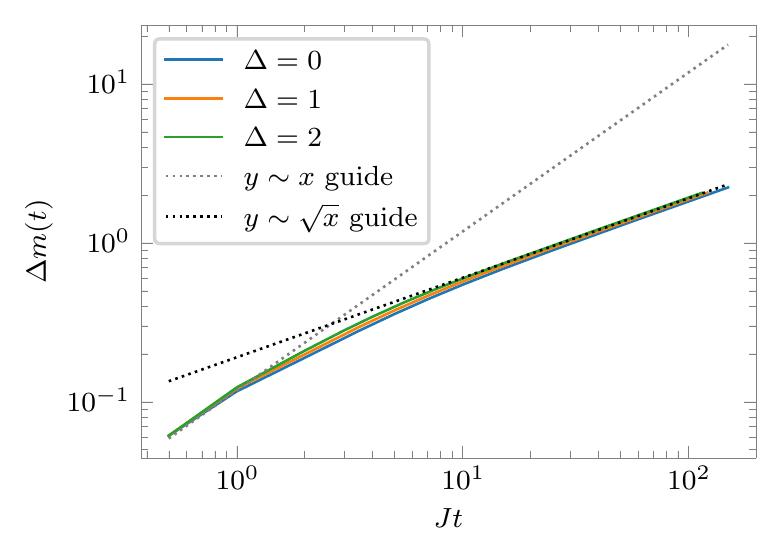}
    \caption{Transference of magnetization across $q=0$ as a function of time. (Dotted lines are guides for the eye.) The bath is the same as the one considered in Fig.~\ref{fig:spin_dyn_D0}.}\label{fig:transfer_mag}
\end{figure}

In Fig.~\ref{fig:transfer_mag}, we plot the amount of magnetization transferred
across $q=0$, $\Delta m$, in presence of the phonons as a function of time for
$\Delta = 0, 1$ and 2. The asymptotic behavior of the system is the same
irrespective of the amount of anisotropy. While the initial dynamics is
ballistic, the dynamics very quickly becomes diffusive.  Notice that the
``super-ballistic'' transients observed in Fig.~\ref{fig:dope_mag} has vanished
and been replaced by an insignificant duration of ballistic transport rapidly
decaying into a uniformly diffusive transport.

\begin{figure}
    \includegraphics{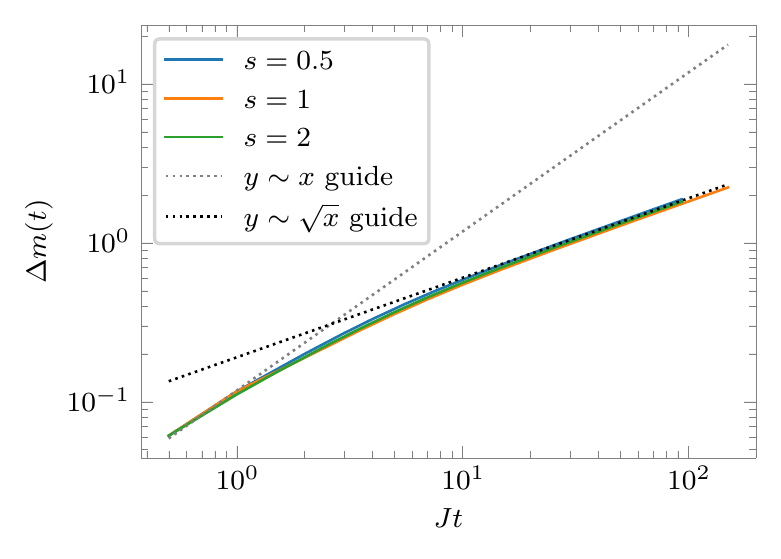}
    \caption{Transference of magnetization across $q=0$ as a function of time in presence of a sub-Ohmic ($s=0.5$), Ohmic ($s=1$) and super-Ohmic($s=2$) baths.}\label{fig:super_sub}
\end{figure}

Till now we have only considered Ohmic baths. Next consider the effect of
different types of bath. Since the biggest change in the nature of the dynamics
happens for the XXZ spin chain with $\Delta = 0$, we explore this effect for
this case. If it is the presence of the bath and not the nature thereof that
causes the diffusive transport, the asymptotic behavior of $\Delta m(t)$ would
scale in the same way, irrespective of the value of $s$.
Fig.~\ref{fig:super_sub} shows the transfer of magnetization across $q=0$ for a
sub-Ohmic ($s=0.5$), Ohmic ($s=1$) and super-Ohmic ($s=2$) baths with $\omega_c
    = 10J$, $\xi = 1$ and $\hbar\omega_c\beta = 5$. We see that irrespective of
whether the bath is Ohmic or sub- or super- Ohmic, the dynamics asymptotically
becomes diffusive.

\begin{figure}
    \includegraphics{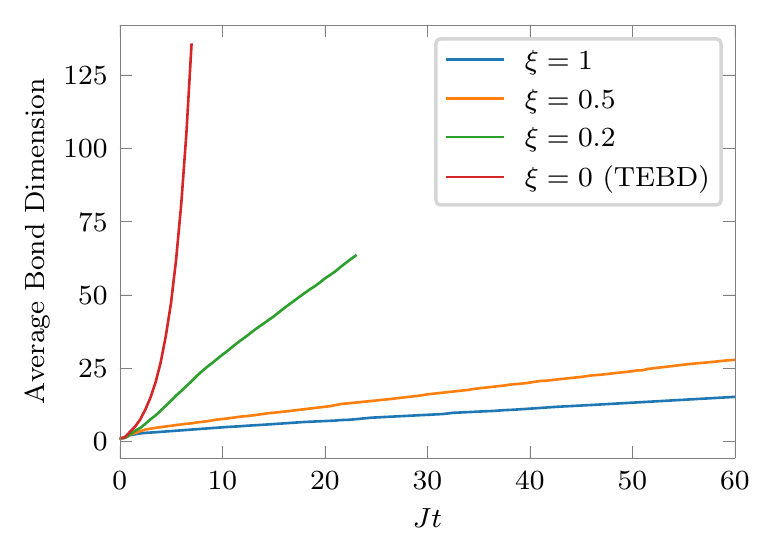}
    \caption{Average bond dimensions of the time propagated reduced density matrix corresponding to the XXZ system with $\Delta = 0$ coupled with different baths.}\label{fig:bond_dims}
\end{figure}

\begin{figure}
    \includegraphics{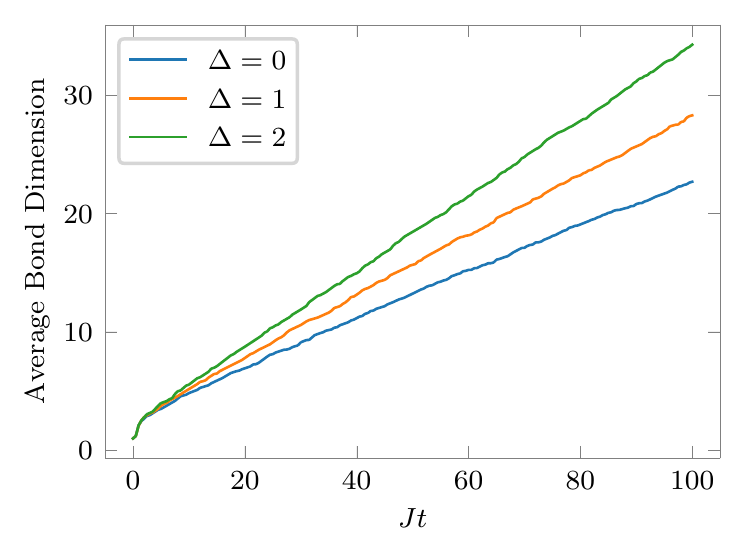}
    \caption{Average bond dimensions of the time propagated reduced density matrix corresponding to the XXZ system with different values of anisotropy when coupled to the Ohmic bath used in Fig.~\ref{fig:spin_dyn_D0}.}\label{fig:bond_dims_anisotropy}
\end{figure}

Finally, we investigate the entanglement in the systems for the different cases.
Instead of explicitly calculating the entanglement entropy, we report the
average bond dimension of the reduced density matrix MPS as an indirect but
simpler measure.  MS-TNPI simulates the reduced density matrix, consequently,
the average bond dimension of the MPS that represents it is going to be larger
than that which represents a wavefunction MPS. Therefore, for this comparison,
we implemented a density matrix version of TEBD. The average bond dimensions of
simulations with different strengths of the phonon bath for the XX system is
shown in Fig.~\ref{fig:bond_dims}. The bond dimension is the smallest for the
strongest coupled bath. The intuition here is that the entanglement goes out
into the bath when it is coupled. The stronger the system-bath coupling, the
more efficient the bath is at limiting the growth of entanglement within the
system. Figure~\ref{fig:bond_dims_anisotropy} shows the growth of the bond
dimension for various values of $\Delta$. The bond dimension grows faster at
higher anisotropy.

\section{Conclusions}\label{sec:conclusion}

Quantum transport in extended systems is a domain of study that is both
incredibly rich both in terms of the physics involved as well as from the
perspective of the potential applications.  The nature of this dynamics is
modified by scattering of the spinons off impurities present in the system and
the phonons with which they couple. These phenomena have been postulated to be
the cause behind the diffusive heat transport observed by
\citet{hlubekSpinonHeatTransport2012}. In this paper, we have numerically
explored the impact of both mechanisms and attempted to clearly attribute the
effects.

Scattering off impurities is relatively easy to simulate using well established
methods like tDMRG and TDVP. \citet{znidaricWeakIntegrabilityBreaking2020} has
studied the effect of periodic on-site magnetic fields. In this paper, we
modeled the impurities as sites with different interactions than the ones in the
clean model with $\Delta = 0$. We demonstrated that the impact of the scattering
events at intermediate timesclaes is highly dependent on the amount and the
nature of the impurities present. However, at long times, the dynamics in
presence of impurities became diffusive. The dynamics was simulated as the
average dynamics over arbitrary configurations of impurities. The nature of the
transport at intermediate times is bounded by the dynamics in the pure system on
one side and the dynamics in a system with 100\% impurities on the other hand.
The time-scale of the diffusive dynamics setting in is dependent on nature of
the impurities.  Future work will explore in further detail the various facets
of dynamics in presence of such impurities.

The coupling to phonons is significantly more challenging to incorporate because
of the presence of non-Markovian memory. Feynman-Vernon influence functional is
often used to simulate the dynamics of open quantum systems. We have recently
developed the multisite tensor network path integral approach to combine ideas
from other tensor network methods like tDMRG and TEBD with influence functional.
This allows us to simulate quantum dynamics of open extended systems without
invoking perturbation theory or Markovian approximations. MS-TNPI is numerically
exact under convergence.

Using MS-TNPI, we analyzed the effect of phonons on the quantum transport. We
demonstrated that irrespective of the system and the description of the phononic
bath, the thermal transport is always diffusive. This is consistent with the
experimental observations in Ref.~\cite{hlubekSpinonHeatTransport2012}. We
hypothesize that this result will carry over to other structured spectral
densities describing the phononic bath. One subtle but notable difference
between the dynamics from the system coupled to phonons and the system with
impurities is the absence of ballistic wavefronts in presence of the phonons.

The presence of the phonons helps ``dissipate'' the growing entanglement of the
XXZ spin chain. We show that the growth of the average bond dimension of the
reduced density matrix corresponding to the extended system is severely
restricted by the presence of the baths. The stronger the coupling to the bath,
the slower the bond dimension grows.

While the fully polarized domain boundary explored here is a special initial
condition in many ways, there have been studies about the dynamics of the XXZ
system with a high temperature density matrix~\cite{ljubotinaSpinDiffusionInhomogeneous2017}, and the so-called
``tilted'' domain boundary~\cite{ljubotinaClassStatesSupporting2017} initial
conditions. These studies explore the rich physics demonstrated by the XXZ
system. It would also be very illuminating to study the transport from such
nonequilibrium initial conditions in presence of the interaction with phonons.
Given the diffusive nature of transport from the fully polarized boundary
initial condition, one may hypothesize that the transport from the other initial
conditions in presence of phonons might also be diffusive. We have recently
studied the effect of temperature gradient on exciton transport in the Frenkel
model~\cite{boseEffectTemperatureGradient2022}. It would be interesting to study
the effect of temperature profile on the diffusion observed from these boundary
initial conditions.

In this work the phonons were coupled only along the Z-direction. In the future,
MS-TNPI will be extended to handle phonon couplings along multiple non-commuting
system operators. It will be interesting to explore the effects of different
couplings. One wonders if the anisotropy of the couplings could be the main
reason behind the uniformly diffusive dynamics observed here.

\section*{Acknowledgments}
I acknowledge the support of the Computational Chemical Sciences Center:
Chemistry in Solution and at Interfaces funded by the US Department of Energy
under Award No.  DE-SC0019394.

\appendix
\section{Transport dynamics in presence of Ohmic bath}\label{app:dynamics}
\begin{figure*}
    \subfloat[Dynamics for $\Delta = 1$.]{\includegraphics{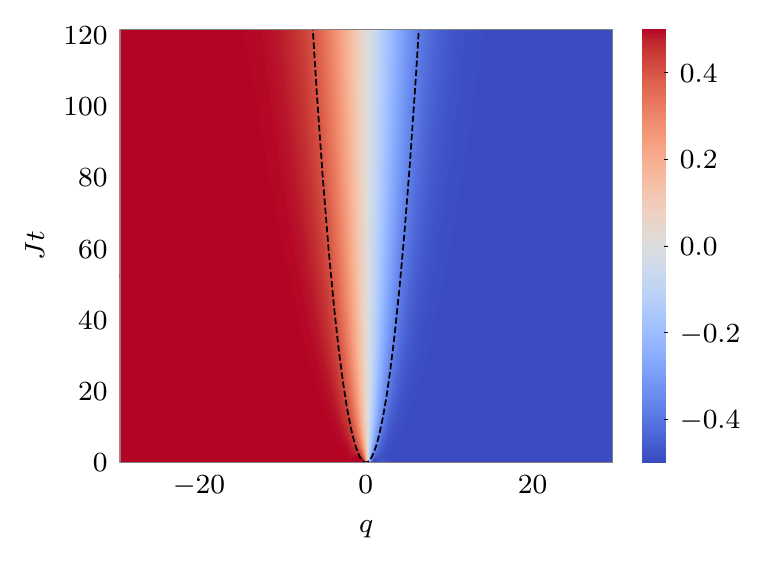}}
    ~\subfloat[Expectation values of $\hat{s}_q^{(3)}$ for $\Delta = 1$ at different time points]{\includegraphics{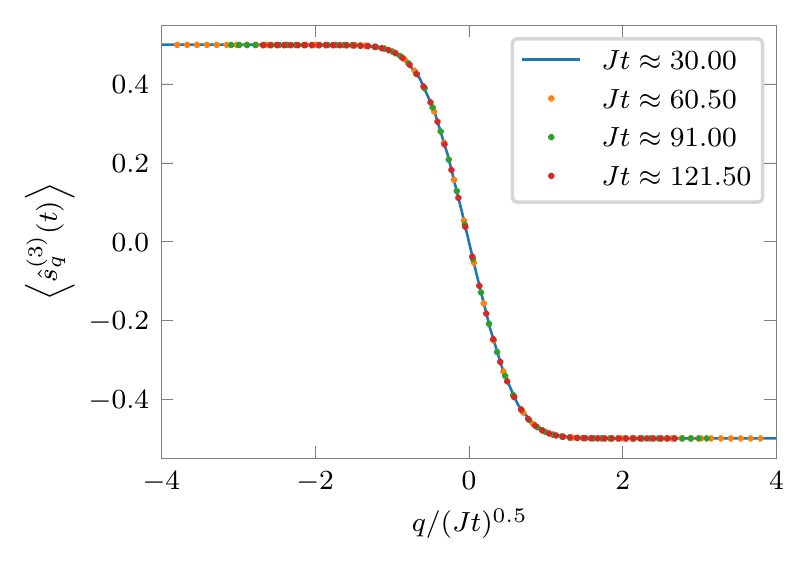}}

    \subfloat[Dynamics for $\Delta = 2$.]{\includegraphics{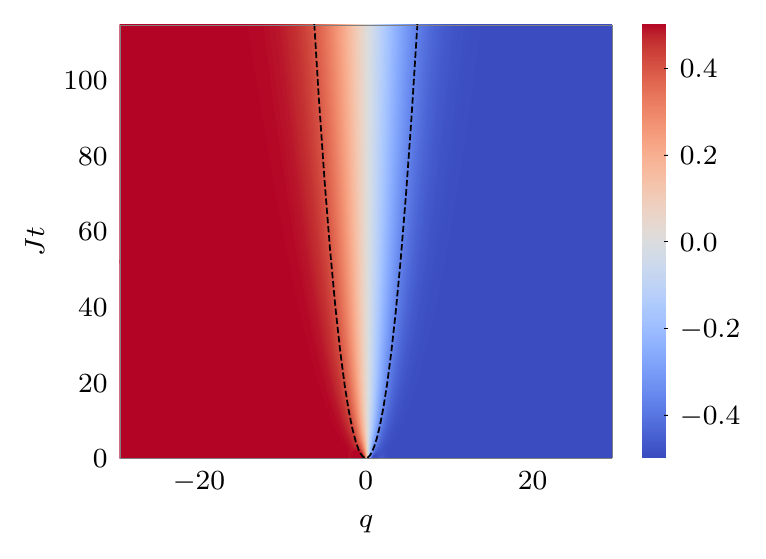}}
    ~\subfloat[Expectation values of $\hat{s}_q^{(3)}$ for $\Delta = 2$ at different time points]{\includegraphics{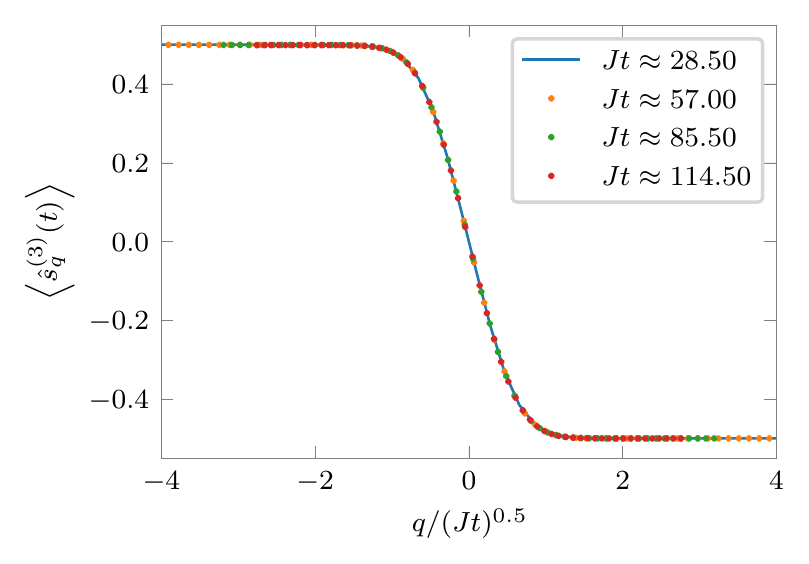}}

    \caption{Dynamics of $\expval{\hat{s}_q^{(3)}(t)}$ and the invariance of the spin profile at different times for XXZ chains with different values of $\Delta$. Left column shows $\expval{\hat{s}_q^{(3)}(t)}$. The black dashed lines guide the eye towards a scaling of $q\sim\sqrt{t}$. Right column shows the spin profiles at different times rescaled by $\alpha=0.5$. The bath used here is identical to the one in Fig.~\ref{fig:spin_dyn_D0}.}\label{fig:spin_dyn_anisotropy}
\end{figure*}

In Sec.~\ref{sec:phonons}, we have shown the dynamics of the XXZ chain with
$\Delta = 0$ connected to a phononic bath described by the Ohmic spectral
density. Here, in Fig.~\ref{fig:spin_dyn_anisotropy}, we show similar graphs for
$\Delta = 1$ and $\Delta = 2$. Note that the dynamics looks identical in all
these cases, implying that the transport process is completely modulated by the
scattering from the phonons.

\bibliography{bibexport.bib}
\end{document}